\documentclass[twocolumn, aps, prl, 10pt,superscriptaddress]{revtex4-1}
\usepackage{graphicx}
\usepackage{amsmath}
\usepackage{amssymb}
\usepackage{hyperref}
\usepackage{color}
\usepackage{ulem}

\begin{document}

\title{Weyl points and topological nodal superfluids in a face-centered
cubic optical lattice}
\author{Li-Jun Lang}
\thanks{They contribute equally to this work.}
\affiliation{Department of Physics, The Chinese University of Hong Kong, Shatin, New Territories, Hong Kong, China}
\author{Shao-Liang Zhang}
\thanks{They contribute equally to this work.}
\affiliation{Department of Physics, The Chinese University of Hong Kong, Shatin, New Territories, Hong Kong, China}
\affiliation{School of Physics, Huazhong University of Science and Technology, Wuhan, China}
\author{K. T. Law}
\affiliation{Department of Physics, Hong Kong University of Science and Technology, Clear Water Bay, Hong Kong, China}
\author{Qi Zhou}
\email{qizhou@phy.cuhk.edu.hk}
\affiliation{Department of Physics, The Chinese University of Hong Kong, Shatin, New Territories, Hong Kong, China}
\affiliation{Department of Physics and Astronomy, Purdue University, West Lafayette, IN, 47906, USA}
\date{\today }

\begin{abstract}
  We point out that a face-centered cubic (FCC) optical lattice, which can be
  realised by a simple scheme using three lasers, provides one a highly
  controllable platform for creating Weyl points and topological nodal
superfluids in ultracold atoms. In non-interacting systems, Weyl points
automatically arise in the Floquet band structure when shaking such FCC
lattices, and sophisticated design of the tunnelling is not required. More
interestingly, in the presence of attractive interaction between two
hyperfine spin states, which experience the same shaken FCC lattice, a
three-dimensional topological nodal superfluid emerges, and Weyl points show
up as the gapless points in the quasiparticle spectrum. One could either create a double Weyl point of charge 2, or
split it to two Weyl points of charge 1, which can be moved in the momentum space by tuning the
interactions. Correspondingly, the Fermi arcs at the surface may be linked with each other or separated as individual ones.
\end{abstract}

\maketitle

Fascinating progresses have been made in studying topological matters of
ultracold atoms in the past few years. A number of topological models and topological phenomena difficult
to access in solid materials have been realised\cite{Aidelsburger2013,Miyake2013,Jotzu2014,Mancini2015,Huang2015}. For
instance, the Harper-Hofstadter model\cite{Hofstadter1976} and
the topological Haldane model\cite{Haldane1988} have been delivered in optical
lattices\cite{Aidelsburger2013,Miyake2013,Jotzu2014,Mancini2015}. In the continuum, a two-dimensional (2D) synthetic spin-orbit coupling
has created a single stable Dirac point that can be moved anywhere in the
momentum space\cite{Huang2015}. So far, most of these studies have been
focusing on one or two dimensions. A large class of three-dimensional (3D)
topological phenomena remain unexplored at the moment.

A Weyl point is a characteristic  3D topological band structure \cite{VOLOVIK2003},
which provides an analog of Weyl fermions, a building block in quantum field theory. It also serves
as an ideal platform to explore a wide range of topological phenomena
in gapless quantum systems, such as Fermi arc\cite{Wan2011} and chiral anomalies\cite{Nielsen1983}.
Weyl points and Weyl semimetals have recently be discovered in certain solid materials\cite{Lv2015,Xu2015,Xu2015a}. Whereas this development represents a major
advancement in the current frontier of condensed matter physics, challenges
remain on manipulating Weyl points, since microscopic parameters are
essentially fixed in a given solid material. Further more, a fundamentally
important question regarding the interplay between Weyl semimetal and
interaction remains unsolved. Though theoretical studies have predicted a
variety of interesting results, including novel superconductivity in doped
Weyl semimetals\cite{Cho2012,Lu2015,Bednik2015,Zhou2015,Li2015,Jian2015,Yuan2016} and the emergent supersymmetry\cite{Jian2015}, there have been no
experimental observation of such phenomena. It is therefore desirable to
have a highly controllable platform to investigate Weyl points and the
resultant quantum phenomena in interacting systems.

\begin{figure}[tbhp]
  \centering
  \includegraphics[width=1\linewidth]{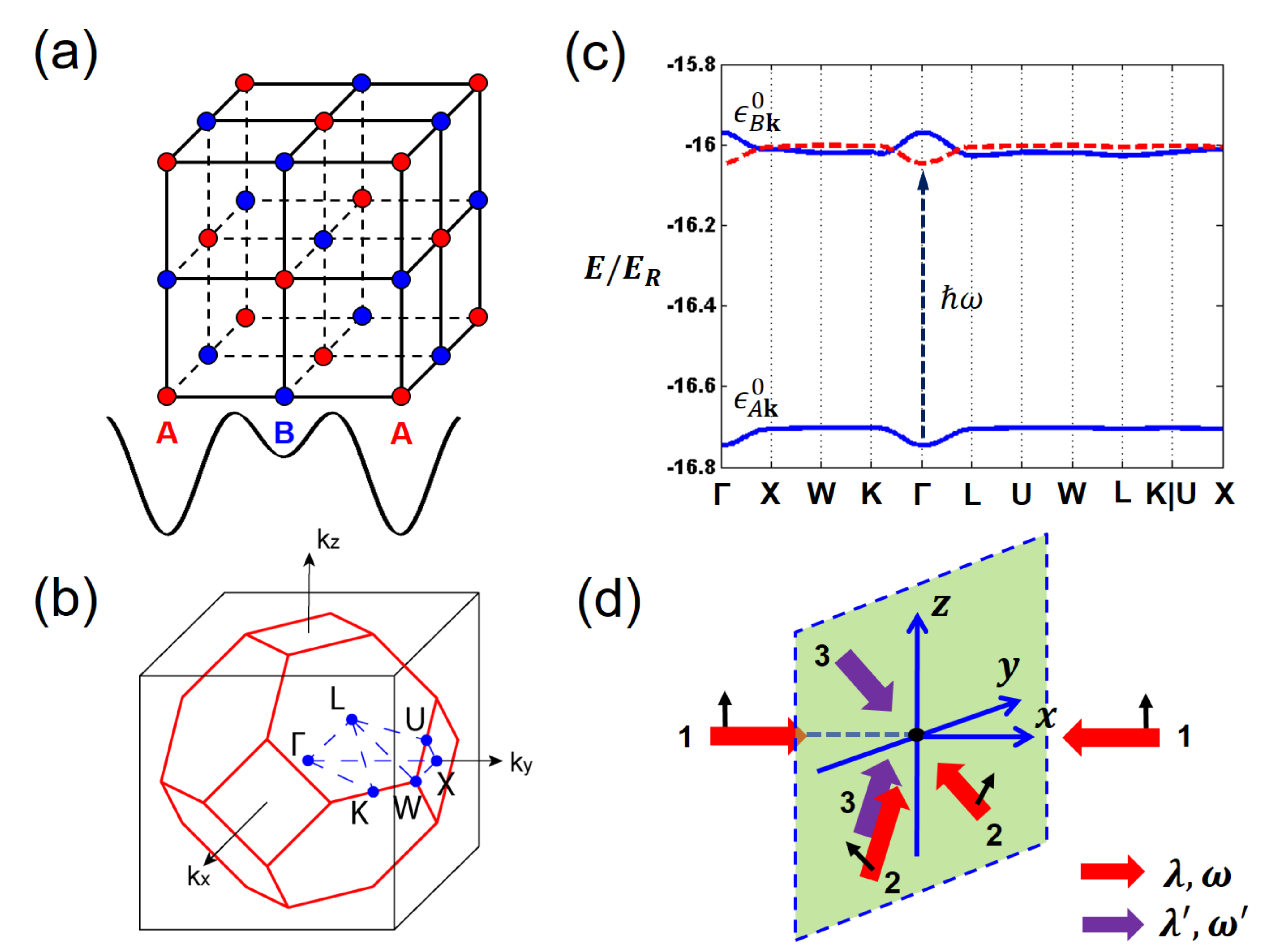}\\
  \caption{(a) Schematic of the FCC lattice with A and B sublattices.
  (b) The 1st BZ (red truncated octahedron) of the FCC lattice folded from that (black cubic) of the SC lattice. (c) The lowest two bands (blue solid), $\epsilon_{A{\bf k}}^0$ and $\epsilon_{B{\bf k}}^0$, of the FCC lattice. Red dashed line represents the subband of $\epsilon_{A{\bf k}}^0$ after absorbing one photon via shaking. The parameters are $V=8E_R, \hbar\omega=0.7E_R, \alpha=0.06, f=0.16d_0$. (d) One laser setup to realize the FCC lattice. Red (purple) arrows represent the lasers with wavelength $\lambda$ ($\lambda'$) and frequency $\omega$ ($\omega'$). Details can be referred to in Supplementary Materials.}\label{lattice}
\end{figure}

In this Letter, we show that a face-centered cubic (FCC) optical lattice
provides physicists a unique means to create and manipulate Weyl points in
both non-interacting and interacting systems. An intrinsic property of a FCC
lattice is that, the lowest two bands, which are labeled as $A$ and $B$, respectively, have ``inverted" band structures, i.e.,
tunnellings with opposite signs. This comes from a simple fact
that the Brillouin zone (BZ) of a FCC lattice is the one folded from a simple cubic (SC) lattice,
as shown in Fig.~\ref{lattice}(a,b). As a result, Weyl points naturally arise in the Floquet band
structure, if one simply uses a periodic shaking to overcome the band
gap and couple the $A$ and $B$ bands. This is distinct from the
majority of previous proposals\cite{Dubfmmodeheckclseciek2015,Xu2015b,He2016,Xu2016}, which require
sophisticated designs to engineer tunnelings along all three directions.
Moreover, uploading two hyperfine spin states onto such optical lattice, the
highly tunable attractive interaction between fermionic atoms allows one to
create a 3D nodal superfluid, which is composed of layered
2D chiral superfluids in the momentum space. Strikingly, Weyl points show up in quasiparticle spectrum of such superfluid. One could either glue two
Weyl points with the same chirality to a monopole of charge 2, or further
split such multiple-charge monopoles into multiple charge-1 ones, and move
them around in the BZ by tuning interactions. Correspondingly, two Fermi arcs emerge at the
surfaces, and can be either linked with or separated from each other,
depending on locations of the Weyl points in the bulk spectrum.

\paragraph{Hamiltonian} We consider a circularly shaken 3D FCC optical lattice,
whose Hamiltonian in the Floquet framework is written as $H_{0}-i\hbar
\partial _{\tau }$, where $\tau $ is time,
\begin{equation}
H_{0}=\frac{\mathbf{P}^{2}}{2M}+V(x+f\cos \omega \tau ,y+f\sin \omega \tau
,z),
\end{equation}%
and $f$ and $\omega $ are the shaken amplitude and frequency, respectively.
$M$ is the atom mass. Formally, the above equation is a direct generalisation of the scheme of shaking a 2D checkerboard lattice\cite{Zhang2015} to a 3D FCC lattice, whose
potential is written as
\begin{eqnarray}
V(x,y,z) &=&-V\big(\cos ^{2}{\frac{\pi }{a}x}+\cos ^{2}{\frac{\pi }{b%
}y}+\cos^{2}{\frac{\pi }{c}z}  \notag \\
&&+\alpha \cos {\frac{\pi }{a}x}\cos {\frac{\pi }{b}y}\cos {%
\frac{\pi }{c}z}\big),  \label{VL}
\end{eqnarray}%
where $a$, $b$ and $c$ are constants.
Interestingly, such a 3D optical lattice of fundamental
importance in solids has never been produced in ultracold atoms. We point
out that it can be produced by three lasers with directions and
polarisations arranged in a few ways, one of which is shown in Fig.~\ref{lattice}(d). Three pairs of lasers are used, where lasers 1 and 2 interfere with each other and form a $z$-dependent checkerboard lattice in the $x$-$y$ plane, while laser 3 with a different frequency forms a standing wave along the $z$ direction (Supplementary Materials). This setup has already been realized by Esslinger's group\cite{Jotzu2014}.

To concretise the discussion, we focus on a symmetric case, $a=b=c$.
All results here can be easily generalised to an arbitrary choice of $%
a,b,c $. The exact band structure of the static lattice $V(x,y,z)$, where $%
f=0$, can be solved exactly using plane-wave expansions. The results for the
lowest two bands, denoted as $\epsilon _{A\mathbf{k}}^{0}$ and $\epsilon _{B%
\mathbf{k}}^{0}$, are shown in Fig.~\ref{lattice}(c). Both of them can be well approximated
by the tight binding results, $\epsilon _{A\mathbf{k}%
}^{0}=4t(\cos k_{x}d_{0}\cos k_{y}d_{0}+\cos k_{y}d_{0}\cos k_{z}d_{0}+\cos
k_{z}d_{0}\cos k_{x}d_{0})+2t^{\prime }(\cos 2k_{x}d_{0}+\cos
2k_{y}d_{0}+\cos 2k_{z}d_{0})$, and $\epsilon _{B\mathbf{k}}^{0}=-\epsilon
_{A\mathbf{k}}^{0}+\Delta$, where $t$ and $t^{\prime }$
characterise the nearest- and next-nearest neighbor tunnelings in the $A$
sublattice, and $\Delta$ is the band gap. The inverted
structure between $A$ and $B$ bands could be understood qualitatively from
the folding of BZ, i.e., $\epsilon _{A\mathbf{k}}^{0}\approx \epsilon _{c%
\mathbf{k}}$, $\epsilon _{B\mathbf{k}}^{0}\approx \epsilon _{c\mathbf{k+G}}+\Delta$
, $\epsilon _{c\mathbf{k}}$ is the ground band dispersion of the SC lattice, and $\mathbf{G}=(1,1,1)\pi/d_0$. $\tilde{d}_0=\sqrt{2}d_{0}$ and $d_{0}$ are the lattice
spacings of the FCC and the SC lattices, respectively. Applying the
shaking, the $A$($B$) band absorbs (emits) a photon and couples with the $B$($A$) band in the Floquet Hamiltonian, which can be written as $H_{0}=(\epsilon
_{A\mathbf{k}}+\epsilon _{B\mathbf{k}})I/2+K$ with
\begin{equation}
K=\left(
\begin{array}{cc}
(\epsilon _{A\mathbf{k}}-\epsilon _{B\mathbf{k}})/2 & \Omega _{\mathbf{k}%
}e^{i\varphi _{\mathbf{k}}} \\
\Omega^*_{\mathbf{k}}e^{-i\varphi _{\mathbf{k}}} & (\epsilon _{B\mathbf{k}%
}-\epsilon _{A\mathbf{k}})/2%
\end{array}%
\right) ,  \label{K}
\end{equation}%
where $\epsilon _{A\mathbf{k}}=\epsilon _{A\mathbf{k}}^{0}+\hbar
\omega$, $\epsilon _{B\mathbf{k}}=\epsilon _{B\mathbf{k}}^{0}$, and $%
\delta =\Delta -\hbar \omega $ is the one-photon detuning. The inter-band coupling is written as $\Omega _{\mathbf{k}%
}e^{i\varphi _{\mathbf{k}}}=\Omega (i\sin k_{x}d_{0}-\sin
k_{y}d_{0})e^{-ik_{x}d_{0}}$ with $\varphi _{\mathbf{k}}=\arg %
\left[ (i\sin k_{x}d_{0}-\sin k_{y}d_{0})e^{-ik_{x}d_{0}}\right]$. This
coupling is the same as that in a shaken checkerboard lattice\cite{Zhang2015}, since $%
V(x,y,z)$ in Eq.(\ref{VL}) reduces to a 2D checkerboard lattice
for each plane with a given value of $z$.

\paragraph{Weyl points from shaking}  A Weyl point requires
that all matrix elements in Eq.~(\ref{K}) become zero at some $\mathbf{k%
}_{0}$ in BZ. In previous proposals\cite{Dubfmmodeheckclseciek2015,Xu2015b,He2016}, this is realised by
engineering the tunnelling along all three dimensions, which often require
sophisticated designs of microscopic models. Here, a shaken FCC lattice
automatically provides one Weyl points. The off-diagonal term could vanish at
$(k_{x},k_{y})=(0,0),(0,\pi ),\text{or~}(\pi ,0)$. Meanwhile, since $\epsilon _{B%
\mathbf{k}}^{0}=-\epsilon _{A\mathbf{k}}^{0}+\Delta$ is
readily satisfied in the static lattice, $\epsilon _{A\mathbf{k}}-\epsilon
_{B\mathbf{k}}=0$ can be easily satisfied if $\delta $ is small enough,
i.e., in the strong inter-band hybridisation regime with the shaken
frequency tuned near resonance. For instance, one may have $\mathbf{k}_{0}%
=(0,0,k_{0z})$ so that $|K|=0$, where $k_{0z}$ satisfies $4t(1+2\cos k_{0z}d_{0})+2t^{\prime }(2+\cos 2k_{0z}d_{0})=\delta /2$. Near
$\mathbf{k}_{0}$, the matrix can be linearised,
\begin{equation}
K=\left(
\begin{array}{cc}
v_{z}(k-k_{0z}) & iv_{x}k_{x}-v_{y}k_{y} \\
-iv_{x}k_{x}-v_{y}k_{y} & -v_{z}(k-k_{0z})%
\end{array}%
\right) ,  \label{KL}
\end{equation}%
where $v_{x}=v_{y}=\Omega d_{0}\equiv v_{\parallel},~v_{z}=-8d_{0}\sin (k_{0z}d_{0})(t+t^{\prime }\cos k_{0z}d_{0})$. Eq.~(\ref{KL}) indeed describes a Weyl point in the BZ. When $\mathbf{k}=\mathbf{k}%
_{0}$, $|K|=0$, and the Weyl point represents a monopole in the momentum
space with a topological charge $\pm 1$. The sign of the charge is
determined by the sign of $v_{x}v_{y}v_{z}$. Since for each $k_{0z}$, there
is always another solution $-k_{0z}$ to satisfy $|K|=0$ with an opposite
chirality, one sees that the total chirality in BZ is zero. The positions of
the Weyl points are determined by the microscopic parameters in the system,
such as the detuning $\delta $. Changing the value of $|\delta |$, Weyl
points move in BZ, and once a pair of Weyl points with opposite
chiralities meet in BZ, they annihilate each other. For large
enough $|\delta |$, the disappearance of Weyl points indicates that the
3D band structure becomes topologically trivial in such cases
with weak inter-band hybridisation.

\begin{figure}[tbhp]
  \centering
  \includegraphics[width=1\linewidth]{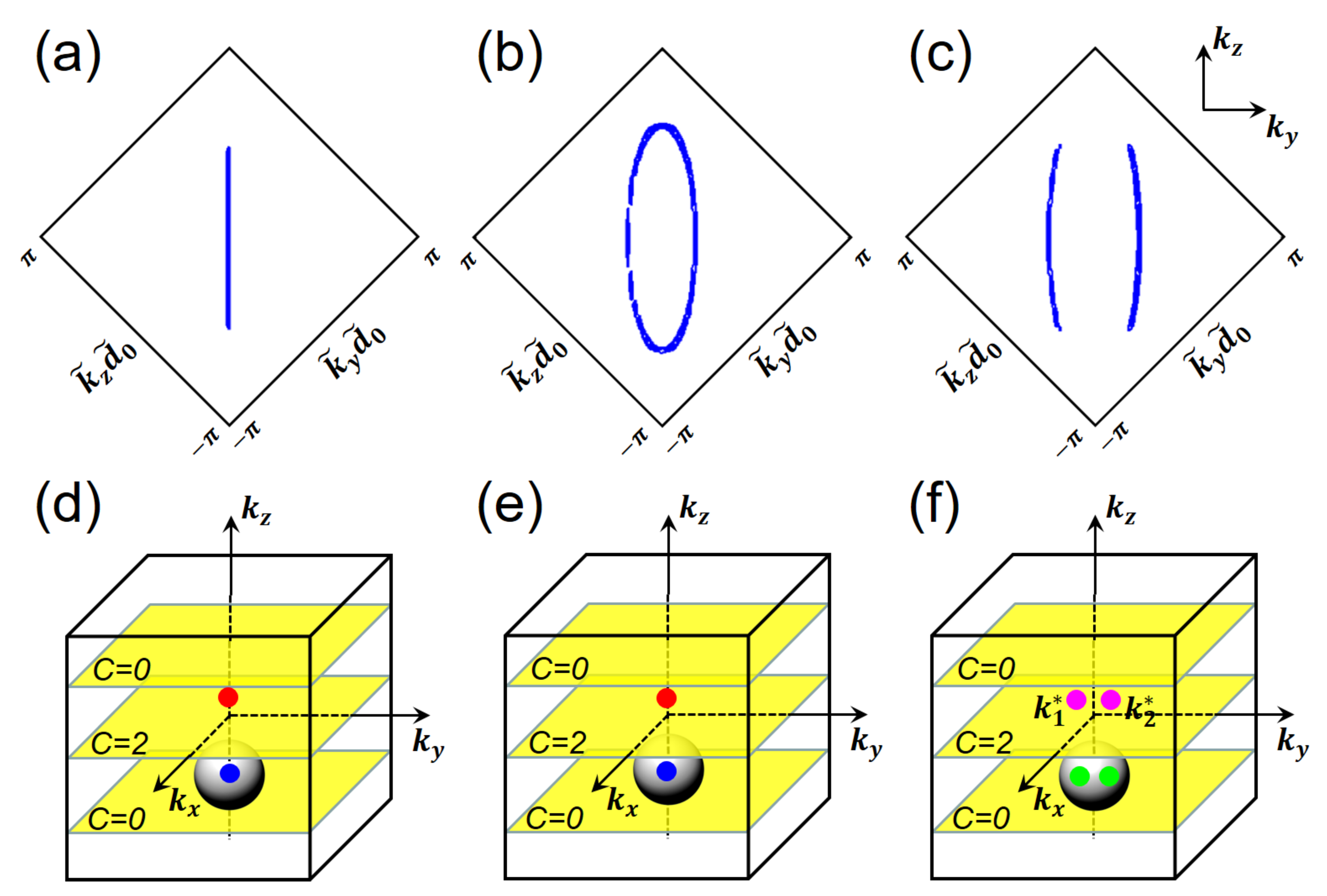}
  \caption{(a-c) Fermi arcs of the single-particle (a) and Bogoliubov quasi-particle (b-c) spectra on the (100) surface of the FCC lattice for half filling with $(U_A,U_B)=(0,0)E_R$, $(U_A,U_B)=(0.15,0)E_R$, and $(U_A,U_B)=(0.15,0.08)E_R$, respectively. $\tilde{k}_{y,z}=(k_z\pm k_y)/\sqrt{2}$. (d-f) Schematics of nodal points in the bulk spectrum corresponding to (a-c), respectively. Red (blue) and pink (green) dots are the nodal points with $+(-)2$ and $+(-)1$ charges, respectively. Spheres enclosing the nodal points have the same Chern number $-2$. Parameters for single particles are the same as those in Fig.~\ref{lattice}(c).} \label{arc}
\end{figure}

\paragraph{3D nodal superfluid}  We now turn to the interaction effects. Whereas the interplay between Weyl
fermions and interaction has been studied in the literature\cite{Cho2012,Lu2015,Bednik2015,Zhou2015,Li2015,Jian2015,Yuan2016}, our shaken
lattices provides one a unique system to explore new physics that has not
been explored before. We introduce two hyperfine spin states into the FCC
optical lattice, each of which has the same single-particle Hamiltonian
specified by Eq.(\ref{K}). This corresponds to a spin-independent shaken
lattice, where we have two Weyl points with the same chirality at the same $%
\mathbf{k}_0$ or $-\mathbf{k}_0$. Without interaction, the
many-body ground state is simply composed of two
identical copies of Weyl semimetals, if the chemical potential is tuned
right at the Weyl point.  Unlike the electronic spins, the hyperfine spin is conserved in
ultracold atoms. This gives rise to a total topological
charge of $2$ in the single particle
level, as shown in fig.~\ref{arc}(d). However, introducing attractive interaction inevitably
leads to particle-hole mixing between the two hyperfine spin states. A
natural question is then, what is the fate of such Weyl semimetals?
Alternatively, one could consider tuning the chemical potential away from Weyl points so that the fermi surface becomes finite. Such finite Fermi surface
indicates that the low lying excitations in the interacting system are
actually located at momenta away from the Weyl points of the non-interacting
system. It is thus desired to explore whether the Weyl point, which is
now embedded inside the Fermi sea, is relevant to the emergent superfluidity when the attractive
interaction is turned on.

Using $\sigma =\uparrow ,\downarrow $ to denote these two hyperfine spin
states, the on-site interaction can be written as $\hat{V}=-U_{A}\sum_{i\in
A}\hat{n}_{i\uparrow }\hat{n}_{i\downarrow }-U_{B}\sum_{i\in B}\hat{n}%
_{i\uparrow }\hat{n}_{i\downarrow }$, where $\hat{n}_{i\sigma }$ is the
density operator for spin-$\sigma $ particles at site $i$, and $U_{A(B)}>0$
is the on-site interaction strength for $A$($B$) sites. Our Hamiltonian is different from the
ones previously studied in the literature, where the degree of freedom that
participates in the interaction is the same as that provides the band
crossing\cite{Bednik2015,Jian2015,Li2015,Lu2015,Zhou2015,Cho2012,Yuan2016}. Here, we consider the attractive interaction between two Weyl
semimetals or two normal metals with small fermi surfaces. Define the pairing
order parameters, $\Delta _{A(B)}=-U_{A(B)}\langle \hat{\Psi}_{A(B),{\bf -k}\downarrow }\hat{\Psi}%
_{A(B){\bf k}\uparrow }\rangle/N $, where $N$ is the number of unit cells, the BCS Hamiltonian can be written as
\begin{equation}
\hat{H}_{BCS}=\left(
\begin{array}{cccc}
\epsilon _{A\mathbf{k}}-\mu & \Omega _{\mathbf{k}}e^{i\varphi _{\mathbf{k}}}
& \Delta _{A} & 0 \\
\Omega _{\mathbf{k}}e^{-i\varphi _{\mathbf{k}}} & \epsilon _{B\mathbf{k}}-\mu
& 0 & \Delta _{B} \\
\Delta _{A}^{\ast } & 0 & \mu -\epsilon _{A\mathbf{k}} & -\Omega _{\mathbf{k}%
}e^{-i\varphi _{-\mathbf{k}}} \\
0 & \Delta _{B}^{\ast } & -\Omega _{\mathbf{k}}e^{i\varphi _{-\mathbf{k}}} &
\mu -\epsilon _{B\mathbf{k}}%
\end{array}%
\right)  \label{HBCS}
\end{equation}%
$\Delta _{A}$ and $\Delta _{B}$ are solved self-consistently for a fixed
density $n=n_{\uparrow }+n_{\downarrow }$. Since $U_{A}$ and $U_{B}$ could
be controlled independently, very rich physics emerges.

We first consider $n_\uparrow=n_\downarrow>1/2$, i.e.,  a finite Fermi surface surrounding each Weyl point in the BZ.
Turning on one of the interactions, say $U_A$ or $U_B$, it turns out that there exists one point on each Fermi surface remaining gapless, whereas the
superfluid gap opens anywhere else, as shown in fig.~\ref{arc}(e). We note that the BCS Hamiltonian~(\ref{HBCS}) is
block-diagonalised if one sets $k_x=k_y=0$. When $U_A=0$, which naturally
leads to $\Delta_A=0$, all matrix elements of one block vanish at the momenta $\mathbf{k}^*=(0,0,\pm k_{0z}\pm k_F)$, i.e.,
the Bogoliubov quasiparticle spectrum remains gapless. The other branch,
which comes from the other block, opens a gap due to a finite $%
\Delta_B$. Alternatively, when $U_B=0$ the spectrum is gapless at $\mathbf{k}^{\prime *}=(0,0,\pm k_{0z}\mp k_F)$.

We point out that the emergent 3D nodal superfluid is a
topological one, which can be seen from the topological charges carried
by the nodal points in the quasi-particle spectrum. Expanding the quasiparticle
spectrum near such gapless point, an effective Hamiltonian is obtained,
\begin{eqnarray}
H_{eff}^{\pm } &=&\left(
\begin{array}{cc}
\pm v_{z}^{\ast }q_{z} & 0 \\
0 & \mp v_{z}^{\ast }q_{z}%
\end{array}%
\right)   \notag \\
&&+\frac{{v}_{\parallel }^{2}(q_{x}^{2}+q_{y}^{2})}{\left\vert \Delta _{\xi
}\right\vert ^{2}+4\mu ^{2}}\left(
\begin{array}{cc}
-2\eta \mu  & \Delta _{\xi}e^{-2i\theta _{q}} \\
\Delta _{\xi }e^{2i\theta _{q}} & 2\eta \mu
\end{array}%
\right) ,
\end{eqnarray}%
where $\mathbf{q}$ is the momentum measured from the gapless point $\mathbf{%
k}^{\ast }$ or $\mathbf{k}^{\prime \ast }$, $\theta _{q}=\arg
(iq_{x}-q_{y})$, and $v_{z}^{\ast }=8\sin (k_{0z}+\eta k_{F})[t+t^{\prime }\cos
(k_{0z}+\eta k_{F})]$. $\eta =+1$ and $\xi=$ A for $U_{B}=0$ while $\eta =-1$ and $\xi=$ B for $U_{A}=0$. The superscript $\pm $ is the valley index, representing
the two Fermi surfaces surrounding the two Weyl points in non-interacting
systems. $H_{eff}^{\pm }$ describes a Bogoliubov quasiparticle spectrum,
which is linear along the $q_{z}$ direction and quadratic along the $q_{x}$
and $q_{y}$ directions, since the off-diagonal term $\sim
(iq_{x}\pm q_{y})^{2}$. It thus describes a monopole of charge-$2$ in the
momentum space. We thus see that the topological charge when introducing attraction interaction
shows up in Bogoliubov quasiparticle spectrum.

Turning on the other interaction, the monopole of charge-$2$ splits to two charge-$1$ ones, i.e., two Weyl points, in the $k_x$-$k_y$ plane, the positions of which are denoted as $\mathbf{k}_1^*$ and $\mathbf{k}_2^*$, respectively. This can be understood from the conservation of the total charge of the monopoles enclosed in the spheres, as shown in fig.~\ref{arc}(e, f). Near these two Weyl points, the dispersion becomes
linear along all three directions. Since the non-interacting system has a
four-fold rotation symmetry about the $k_z$ axis, and such a splitting
reduces the symmetry to a two-fold one, the choice of the direction for the
splitting is a consequence of spontaneous symmetry breaking, along either $%
k_x$ or $k_y$ direction. By changing the ratio $%
U_A/U_B$, these two Weyl points move in BZ before meeting their counterparts
with opposite chiralities emerged from the other valley. For non-interacting
systems, it is known that tuning the parameters leads to the movement of
Weyl points without opening the gap. In our system, tuning interaction also
offers such opportunity to control the positions of Weyl points in the
Bogoliubov quasiparticle spectrum in BZ.

There is an alternative way to understand why the 3D superfluid remains nodal with turning on a small $U_A$. As shown in eq.~\ref{HBCS} and fig.~\ref{arc}(d-f), in 3D BZ each 2D plane with fixed $k_z$ defines an 2D $s+(d+id)$ superfluid, similar to the one emerged from a 2D shaken lattice~\cite{Zhang2015}. Whereas tuning $k_z$ effectively changes the chemical potential of such 2D superfluid, its Chern number can be computed straightforwardly. When $|k_z|<k_z^*$, where $k_z^*$ is the $z$ component of $\mathbf{k}_1^*$ (and $%
\mathbf{k}_2^*$), this 2D chiral superfluid is topologically nontrivial
with a Chern number of 2, as the chiral component $(d+id)$ is dominant for small $U_A$. When $|k_z|>k_z^*$, $C=0$.
The topological transition of the 2D superfluid just corresponds to the nodal points. Thus, the nodal points cannot be suddenly gapped, and the 3D superfluid remains nodal when $U_A$ is small enough.

The pairing here is an inter-valley BCS pairing between $-\mathbf{k}$ and $\mathbf{k}$.
There have been studies of the competition
between the inter- and intra-valley parings near Weyl
points\cite{Cho2012,Lu2015,Bednik2015,Zhou2015,Li2015,Jian2015}. Here, it is the hyperfine spin states that incorporate the interaction effect. We find out
that, at least in the mean field level, the inter-valley pairing wins.
This can be qualitatively understood from the phase space argument.
In the inter-valley pairing, a paired state $({\bf k}_0+{\bf q}\uparrow, -{\bf k}_0-{\bf q}\downarrow)$ can be scattered to both $({\bf k}_0+{\bf q}'\uparrow, -{\bf k}_0-{\bf q}'\downarrow )$ and $(-{\bf k}_0+{\bf q}'\uparrow, {\bf k}_0-{\bf q}'\downarrow )$. For the intra-valley
paring, a paired state $({\bf k}_0+{\bf q}\uparrow, {\bf k}_0-{\bf q}\downarrow )$ could only be scattered to $({\bf k}_0+{\bf q}'\uparrow, {\bf k}_0-{\bf q}'\downarrow )$. Thus, the inter-valley
paring gains more energy than the intra-valley paring (Supplementary Materials).  Another mechanism favoring the inter-valley pairing is that $\epsilon_{\mathbf{k}\sigma}=\epsilon_{-\mathbf{k}\sigma}$ is satisfied. Since the shaking is spin-independent, $\epsilon_{{\bf k}_0+{\bf q}\uparrow}=\epsilon_{ -{\bf k}_0-{\bf q}\downarrow}$ is thus valid. However, near the same Weyl point, $\epsilon_{\mathbf{k_0+q}}$ is not exactly the same as $\epsilon_{\mathbf{k_0}-\mathbf{q}}$, when $\mathbf{q}$ is large enough and the linear
approximation for the single particle energy is no longer accurate. This
energy mismatch disfavours the intra-valley pairing.

We now consider half filling $n=1/2$. In such Weyl semimetal, due to the
vanishing Fermi surfaces, a weak attractive
interaction is no longer relevant\cite{Jian2015}. Our calculations indeed show that
the pairing remains vanishing before either or both interactions reach a critical
value. The critical interaction strength by itself is not
universal, in the sense that it depends on the details of the
single-particle spectrum at both low and high energies. For a
purely linear dispersion, the critical value is written as $U^c_{A}=6\pi^2v_z\Omega^2/\Lambda^2$ for $U_B=0$, where $\Lambda$ is the high energy cutoff. In realistic systems, the high energy part of the spectrum is no longer linear, and is also important to the critical interaction. Thus, there is no simple expression
of $U_A^c$. Nevertheless, if $U_A>U_A^c$ , the results become
similar to those with a finite Fermi surface. A double Weyl point of charge 2 emerges in the
quasi-particle spectrum. Turning on $U_B$, such
multiple-charge monopole splits to two Weyl points of 1.

\paragraph{Fermi arcs} A characteristic feature of Weyl points in non-interacting system is the
existence of Fermi arc\cite{Wan2011} at the surface, an unclosed line as the zero energy
state. Whereas in the absence of interaction, such Fermi arc is indeed
observed in our system as shown in fig.~\ref{arc}(a), it is more interesting to explore the interacting case when the nodal superfluids have emerged. We solve the BDG equation for half filling self-consistently with (100) surface such that $k_y$ and $k_z$ are still good quantum numbers, and the
zero energy surface state is shown in Fig.~\ref{arc}(a-c).

Without interactions, two identical Fermi arcs, each of which comes from one hyperfine spin state, are
on top of each other. Turning on one of the interactions, two Fermi arcs
connect with each other at two points in the 2D BZ, which are
just the projection of the two charge-2 monopoles in the bulk spectrum onto the surface.
Turning on the other interaction, accompanied with the splitting of each
charge-2 monopole into two Weyl points, the two Fermi arcs split, and the
four ending points are simply the projections of the four Weyl points in the
bulk. One could understand these fermi arcs from the layered 2D superfluids. When the 2D superfluid is topologically nontrivial with Chern number 2, its two zero-energy edge states have momenta $k_{y1}(k_z)$ and $k_{y2}(k_z)$, respectively. Changing $k_z$ leads to different values of $k_{y1}$ and $k_{y2}$, and gives rise to
Fermi arcs as the trajectories of  zero-energy states on the surface. Such zero energy states merge into the
bulk spectrum when $|k_z|> k_z^*$ and the 2D superfluid becomes the topologically trivial.

We have shown that Weyl points are readily achievable in current ultracold
atom experiments. The interplay between the interaction and the topological band
structure leads to a 3D topological nodal superfluid with Weyl points and Fermi arcs, which are highly controllable via tuning the interaction. We hope that our work will
stimulate more studies on 3D topological matters in ultracold
atoms.

\begin{acknowledgments}
This work is supported by Hong Kong Research Grants Council/Collaborative Research Fund HKUST3/CRF/13G. Q.Z. acknowledges useful discussions with T. Esslinger.
\end{acknowledgments}

%\bibliographystyle{apsrev4-1}
%\bibliography{ref}


\clearpage

\onecolumngrid

\appendix*

\part{\bf\large Supplemental Material for ``Weyl points and topological nodal superfluids in a face-centered
cubic optical lattice"}

In this supplementary material, we present the results on the realization of the FCC lattices and the comparison between inter- and intra-valley pairings.

\section{Realization of the FCC lattices}

\begin{figure}[tbhp]
  \centering
  \includegraphics[width=0.8\linewidth]{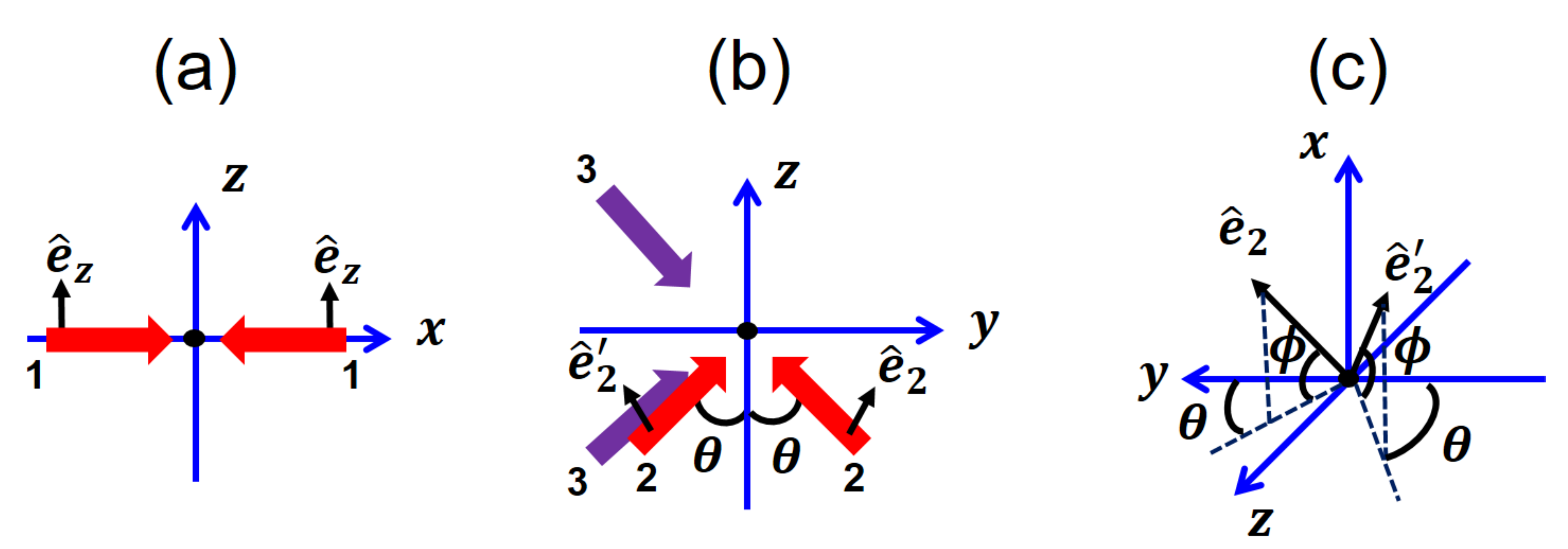}
  \caption{(a) Schematic of laser 1 beams (red arrows) along $x$ axis with linear polarisations ${\bf\hat{e}_z}$ along $z$ direction. (b)Schematic of laser 2 (red arrows) and laser 3 (purple arrows) beams, which are both in $y$-$z$ plane. The linear polarisations, ${\bf\hat{e}_2}$ and ${\bf\hat{e}_2^\prime}$, of laser 2 beams normal to the propagating directions can be in general out of the plane. (c) The polarisation directions, ${\bf\hat{e}_2}$ and ${\bf\hat{e}_2^\prime}$, of laser 2 beams are parameterized by angles $\theta$ and $\phi$.} \label{setup}
\end{figure}

We can use three retro-reflected lasers with linear polarisations to
generate the potential of the face-centered cubic (FCC) lattice. As shown in
Fig. \ref{setup}(a,b), laser 1 in $x$ direction and laser 2 in $y$-$z$ plane both
with linear polarisations have the same wavelength $\lambda _{1}$ and
frequency $\omega _{1}$, and thus can interfere with each other. The total
electric field is%
\begin{eqnarray}
\mathbf{E}&=&2E_{1}e^{i\left( \varphi _{1}+\varphi _{1}^{\prime }\right)
/2}\cos \left( k_{1}x+\frac{\varphi _{1}-\varphi _{1}^{\prime }}{2}\right)
\mathbf{e}_{z}\notag \\
&&+2E_{2}e^{i\left[ k_{1}z\cos \theta +\left( \varphi
_{2}+\varphi _{2}^{\prime }\right) /2\right] }\left( e^{i\left[ k_{1}y\sin
\theta +\left( \varphi _{2}-\varphi _{2}^{\prime }\right) /2\right] }\mathbf{%
e}_{2}+e^{-i\left[ k_{1}y\sin \theta +\left( \varphi _{2}-\varphi
_{2}^{\prime }\right) /2\right] }\mathbf{e}_{2}^{\prime }\right)
\end{eqnarray}%
where $k_{1}=2\pi /\lambda _{1}$ is the wavenumber, $E_{1}$ and $E_{2}$ are
the strengths of the electric fields of lasers 1 and 2, respectively. As
shown in Fig. \ref{setup}(b,c), $\theta$ is the angle of
the direction of laser 2 to the positive $z$ direction, and $\phi$ is the angle of the porlarisation of laser 2 to the $y$-$z$ plane. $\mathbf{e}_{2}=\mathbf{e}_{x}\sin \phi +\mathbf{e}_{y}\cos \phi
\cos \theta +\mathbf{e}_{z}\cos \phi \sin \theta $ and $\mathbf{e}%
_{2}^{\prime }=\mathbf{e}_{x}\sin \phi -\mathbf{e}_{y}\cos \phi \cos \theta +%
\mathbf{e}_{z}\cos \phi \sin \theta $ are the unit vectors of the
polarisation directions of the two laser 2 beams, respectively. $\varphi
_{1(2)}$ and $\varphi _{1(2)}^{\prime }$ are the phases of the incedent and
reflected beams of laser 1(2), respectively. So the corresponding potential
is%
\begin{eqnarray}
V_{12} &=&-V_{1}\cos ^{2}\left( k_{1}x+\frac{\varphi _{1}-\varphi
_{1}^{\prime }}{2}\right) -V_{2}\left( 1-2\cos ^{2}\theta \cos ^{2}\phi
\right) \cos ^{2}\left( k_{1}y\sin \theta +\frac{\varphi _{2}-\varphi
_{2}^{\prime }}{2}\right) \\
&&-2\sqrt{V_{1}V_{2}}\cos \phi \sin \theta \cos \left( k_{1}x+\frac{\varphi
_{1}-\varphi _{1}^{\prime }}{2}\right) \cos \left( k_{1}y\sin \theta +\frac{%
\varphi _{2}-\varphi _{2}^{\prime }}{2}\right) \cos \left( k_{1}z\cos \theta
-\frac{\varphi _{1}+\varphi _{1}^{\prime }-\varphi _{2}-\varphi _{2}^{\prime
}}{2}\right) , \notag
\end{eqnarray}%
where $V_{1,2}$ are the single-beam lattice depths of laser 1 and 2,
respectively. Laser 3 also in $y$-$z$ plane with wavelength $\lambda _{3}$ and
frequency $\omega _{3}$ can form an independent standing-wave potential along $z$ direction, as
long as we make $\left\vert \omega _{3}-\omega _{1}\right\vert \gg \tau $,
where $\tau $ is the typical measurement time, such that it doesn't interfere with other lasers. The potential is%
\begin{equation}
V_{z}=-V_{3}\cos ^{2}\left( k_{3}z+\frac{\varphi _{3}-\varphi _{3}^{\prime }%
}{2}\right) ,
\end{equation}%
where $k_{3}=2\pi\cos\theta/\lambda_3$ is the $z$-component wavenumber of the laser 3, $\varphi _{3}$ and $%
\varphi _{3}^{\prime }$ are the phases of the incedent and reflected beams
of laser 3, respectively, and $V_{3}$ is the single-beam lattice depth. For
convenience, we can tune the parameters as $\varphi _{1}=\varphi
_{1}^{\prime }=\varphi _{2}=\varphi _{2}^{\prime }$ and$\ \varphi
_{3}=\varphi _{3}^{\prime }$, so the potential is reduced to
\begin{eqnarray}
V(\mathbf{r}) &=&-V_{1}\cos ^{2}k_{1}x-V_{2}\left( 1-2\cos ^{2}\phi \cos
^{2}\theta \right) \cos ^{2}\left( k_{1}y\sin \theta \right) -V_{3}\cos
^{2}k_{3}z \notag \\
&&-2\sqrt{V_{1}V_{2}}\cos \phi \sin \theta \cos k_{1}x\cos \left( k_{1}y\sin
\theta \right) \cos \left( k_{1}z\cos \theta \right).
\end{eqnarray}%

If we make $V_{1}=V_{2}\left( 1-2\cos ^{2}\phi \sin ^{2}\theta \right)
=V_{3}\equiv V,$ $k_{1}\cos \theta =k_{3}$, and define $k_{2}\equiv
k_{1}\sin \theta $ and $\alpha \equiv 2\cos \phi \sin \theta \sqrt{V_{1}V_{2}%
}/V$. The lattice potential becomes%
\begin{equation}
V(\mathbf{r})=-V\left( \cos ^{2}k_{1}x+\cos ^{2}k_{2}y+\cos
^{2}k_{3}z+\alpha \cos k_{1}x\cos k_{2}y\cos k_{3}z\right).
\end{equation}%
This is just the potential for FCC lattic

\section{Comparison between inter- and intra-valley pairings}

\begin{figure}[tbhp]
  \centering
  \includegraphics[width=1\linewidth]{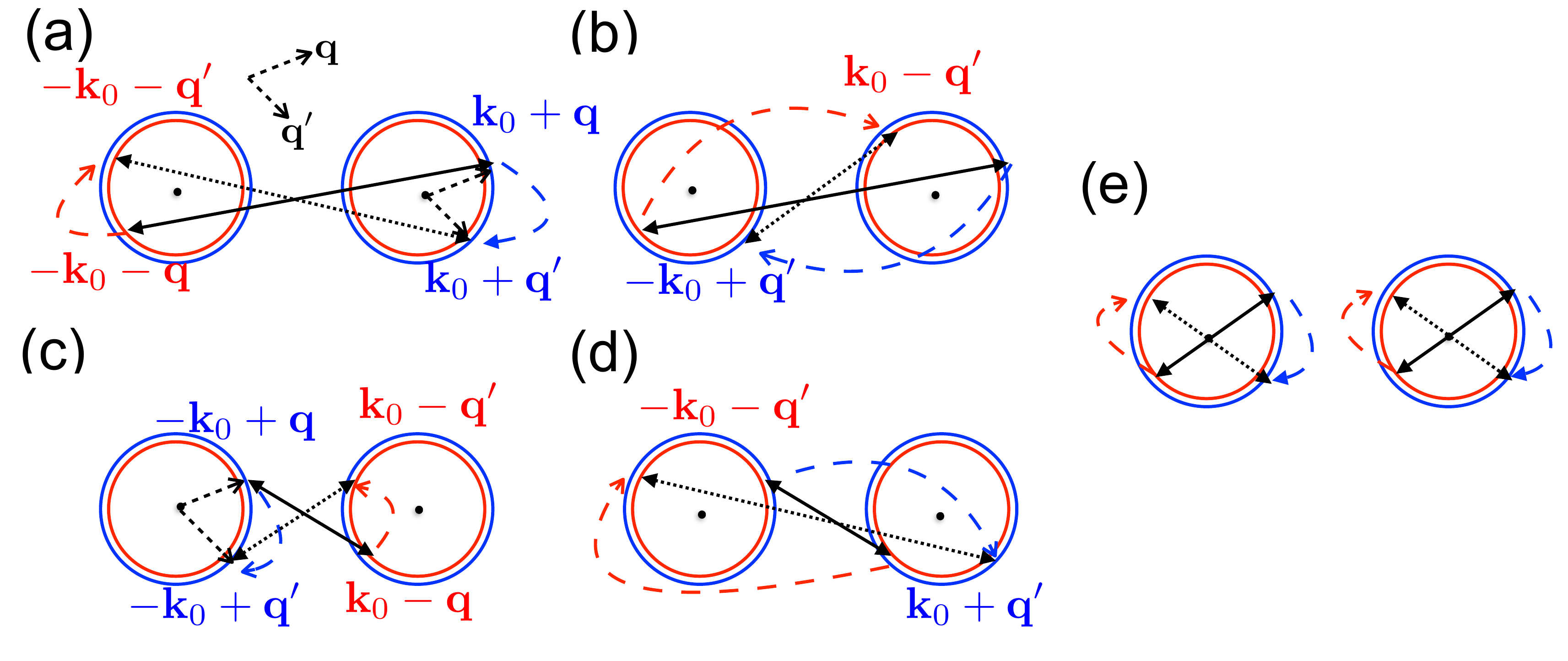}
  \caption{Blue and red circles represent the identical Fermi surfaces of spin-up and spin-down atoms, respectively, which have been slightly displaced for clarity. Solid and dotted arrows represent the initial and final momenta, respectively. Dashed curves highlight the changes of the momentum of spin-up (blue) and spin-down (red) atoms.  (a-d) Inter-valley paring. (e) Intra-valley pairing.  } \label{valley}
\end{figure}

Consider the inter-valley pairing, for any fixed ${\bf q}$ and  ${\bf q}'$, the initial state of a pair $({\bf k}_0+{\bf q}, \uparrow; -{\bf k}_0-{\bf q}, \downarrow)$ can be scattered to $({\bf k}_0+{\bf q}', \uparrow; -{\bf k}_0-{\bf q}', \downarrow)$, as shown in Fig (\ref{valley}a). For the same values of ${\bf q}$ and  ${\bf q}'$, the same initial state can also be scattered to  $(-{\bf k}_0+{\bf q}', \uparrow; {\bf k}_0-{\bf q}', \downarrow)$, which also conserves the total momentum and energy, as shown in Fig (\ref{valley}b). Similarly, the initial state of a pair $(-{\bf k}_0+{\bf q}, \uparrow; {\bf k}_0-{\bf q}, \downarrow)$ can be scattered to $(-{\bf k}_0+{\bf q}', \uparrow; {\bf k}_0-{\bf q}', \downarrow)$ in Fig (\ref{valley}c) or $({\bf k}_0+{\bf q}', \uparrow; -{\bf k}_0-{\bf q}', \downarrow)$ in Fig (\ref{valley}d). Thus the energy gained by the inter-valley pairing is proportional to $\int d{\bf q}d{\bf q}'\sim 4 \mathcal{N}(E_F)$, where $\mathcal{N}(E_F)$ is Density of States of a single Fermi surface of a single spin component. In contrast, for intra-valley pairing, the initial state of a pair $(\pm{\bf k}_0+{\bf q}, \uparrow; \pm{\bf k}_0-{\bf q}, \downarrow)$ can be scattered to $(\pm{\bf k}_0+{\bf q}', \uparrow; \pm{\bf k}_0-{\bf q}', \downarrow)$ in the same valley only, as shown in Fig (\ref{valley}e). Thus the energy gained by the intra-valley pairing is  proportional to $\int d{\bf q}d{\bf q}'\sim 2 \mathcal{N}(E_F)$. One thus conclude inter-valley pairing is in general favored in our system.

\end{document}